\documentclass[11pt]{article}

\usepackage{graphicx} 
\usepackage[margin=1in]{geometry} 
\graphicspath{{Figures/}} 
\usepackage{hyperref} 
\hypersetup{colorlinks}
\usepackage{amsmath} 
\usepackage{amssymb} 
\usepackage{multirow} 
\usepackage{booktabs} 
\usepackage{color} 
\usepackage{indentfirst} 
\usepackage{caption, subcaption} 
\usepackage{bm} 
\usepackage{siunitx} 
\usepackage[noblocks]{authblk} 
\usepackage{microtype} 

\newcommand{\im}{\mathrm{i}}

\begin{document}
	
\newcommand{\beginsupplement}{%
		\setcounter{table}{0}
		\setcounter{figure}{0}
		\setcounter{section}{0}
		\setcounter{subsection}{0}
		\setcounter{equation}{0}
		\setcounter{paragraph}{0}
		\setcounter{page}{1}
		\renewcommand{\thetable}{S\arabic{table}}%
		\renewcommand{\thefigure}{S\arabic{figure}}%
		\renewcommand{\thesection}{S\arabic{section}}%
		\renewcommand{\thesubsection}{S\arabic{section}.\arabic{subsection}}%
		\renewcommand{\theequation}{S\arabic{equation}}%
		\renewcommand{\theparagraph}{S\arabic{paragraph}}%
		\renewcommand{\thepage}{S\arabic{page}}%
		\renewcommand{\theHtable}{S\the\value{table}}%
		\renewcommand{\theHfigure}{S\the\value{figure}}%
		\renewcommand{\theHsection}{S\the\value{section}}%
		\renewcommand{\theHsubsection}{S\the\value{subsection}}%
		\renewcommand{\theHequation}{S\the\value{equation}}%
		\renewcommand{\theHparagraph}{S\the\value{paragraph}}%
}
	
\title{Optimization of Graded Metamaterials for Control of Energy Transmission Using a Genetic Algorithm}
\author[1]{Joshua Morris}
\author[1]{Weidi Wang}
\author[2]{Thomas Plaisted}
\author[1]{Christopher J. Hansen}
\author[1,3]{Alireza V. Amirkhizi}
\affil[1]{Mechanical Engineering, University of Massachusetts Lowell, Lowell, MA USA}
\affil[2]{DEVCOM ARL, Aberdeen Proving Ground, MD, USA}
\affil[3]{alireza\_amirkhizi@uml.edu}
\date{\today}
\maketitle

\renewenvironment{abstract}
{\begin{quote}
        \noindent \rule{\linewidth}{.5pt}\par{\bfseries \abstractname.}}
    {\medskip\noindent \rule{\linewidth}{.5pt}
\end{quote}}
\begin{abstract}
Optimization of functionally graded metamaterial arrays with a high dimensional and continuous geometric design space is cumbersome and could be accelerated via machine learning tools. Mechanical metamaterials can manipulate acoustic or ultrasonic waves by introducing large dispersive and attenuative effects near their natural frequency. In this work functionally graded structures are designed and optimized to combine the energy attenuation performance of a number of unit cells with varying frequency responses and to reduce the interlayer mismatch effects. Optimization through genetic algorithm avoids the many local minima related to high dimensionality of the design space but requires many iterations. A reduced order model (ROM) is applied that can reproduce the transmission response that is traditionally calculated with FEM, in a fraction of the time. Pairing GA and the ROM together, an array of 6 unit cells (with a total of 18 independent geometric  design variables) is optimized to have stop bands with extended width and sharper boundaries. Symmetric functionally graded structures are determined to be optimal geometric configurations. Measured 3D printed features are projected onto the ROM solutions to quantify the effect of printing uncertainty on array performance. Repeatability error of $\pm~20$ $\mu$m is determined to reduce the mean depth of the transmission stop band by a factor of $10^2$ and introduce small shifts in center frequency and band width. Proposed methods to improve the resolution of accessible points in the ROM space, reduce sensitivity to geometric uncertainty, and add design freedom include introducing out-of-plane perforations and varying constituent materials using tunable filled resin systems.  \\
\end{abstract}


\maketitle

\section{Introduction}

Mechanical metamaterials have highly tunable performance and display engineered dynamic properties not observed in traditional materials. When transitioning from simulated designs to practical implementation, the large number of design space variables presents a daunting optimization problem. Machine learning tools have enabled the tailoring of metamaterial systems for specific applications, such as sound insulation \cite{zhao_machine-learning_2021}, acoustic lenses \cite{li_design_2012}, and cloaking \cite{pomot_acoustic_2020,meng_optimization_2012}. Model selection is highly dependent on the characteristics of the system. Gaussian processes coupled with Johnson–Champoux–Allard \cite{casaburo_gaussian-based_2022}, Bayesian \cite{zheng_inverse_2020}, or Radial Basis Functions \cite{bacigalupo_machine-learning_2020} are some of the highly regarded methods for continuous systems and are well suited for regression approaches with small training data sets. Similarly, topology optimization via neural networks \cite{he_inverse_2021,wu_machine_2020,kollmann_deep_2020,wu_design_2021,gazonas_genetic_2006} has been popular as a method of reducing the number of required iterations to solve highly parametric (often discontinuous) problems. On the other hand, genetic algorithm (GA) requires a very large number of iterations compared to its competitors, but could avoid local minima that develop in systems with large dimensionality. Wang and Liu \cite{wang_parameter_2021} integrated GA with finite element method (FEM) to optimize the efficiency of their bidirectional re-entrant honeycomb (auxetic) structure and, while successful, needed to limit the number of iterations using a control variable to reduce the computational cost. Wu et al. \cite{wu_parametric_2022} utilized GA to maximize band gap width for multiple aperiodic unit cells, similar to the present work. Their approach applies Bloch wave equations to produce a reduced order solution that decreases the computational cost of each GA step, and was successfully used to expand their band gap considerably at extremely low frequency ranges (10-25 Hz) using an objective function linked to the imaginary part of the wave number. To avoid pitfalls with the computational cost and maintain design freedom, a reduced order model (ROM) or closed form formulation is almost necessary when leveraging GA to find global minima in problems with large dimensionality. 

Optimization of finite arrays, as compared to infinite periodic media, frequently steers efforts toward the development of functionally graded systems. In an early success, Cummer and Schurig \cite{cummer_one_2007} analytically proved that a metamaterial layer with continuously graded density could theoretically achieve  cloaking. In reality, continuous gradients generally need to be broken down into a series of discrete, imperfect steps \cite{craster_long-wave_2014,he_multi-objective_2018,amirkhizi_continuous_2017}. Functionally graded materials can be organized into three classes \cite{zhang_additive_2018,li_review_2020}: geometric variation with a controlled material \cite{sepehri_tunable_2020}, material variation with a controlled geometry \cite{kuang_grayscale_2019}, or a combination of both \cite{jafari_hybrid_2020}. Romero-Garcia et al. \cite{romero-garcia_design_2021} take a similar geometric approach as seen in the present work, where they produce a short array of 9 unique resonator cells each with slightly shifted stop bands.
Their work utilizes Helmholtz resonators to maximize absorption over a 200-1200 Hz range as they manually tailored the response of their system and experimentally validated the broad band absorption.

This work is inspired by the successes of leveraging reduced order models for use in genetic algorithm as a means of optimizing transmission loss for a short array of locally resonant unit cells. First, an expansion of a Bloch-Floquet based ROM to approximate 1D transmission is presented, including the reduced order bounds for the cell geometry. Genetic algorithm is then applied for optimization of the geometry for 6 unique unit cells (for a total of 24 ROM parameters based on 18 independent geometric variables) with an objective function that minimizes specific integrals associated with the transmission curve. 
The optimization tool identifies arrays with gradient properties or symmetry as providing performance benefits compared to a series of repeating cells with uniform geometry. Finally, the manufacturing constraints that limit the resolution of the functionally graded array are studied. The process for their consideration involves assigning uncertainty regions around a desired design value and analyzing the effect of such uncertainties on the performance metric. Creative design options are presented as a means to reduce sensitivity to manufacturing error and improve the performance of cellular architectures.

\section{Reduced Order Model (ROM) for Transmission}

\begin{figure*}[!htb]
  \centering
  \begin{subfigure}[htb]{0.35\textwidth}
    \centering
    \includegraphics[scale=0.2]{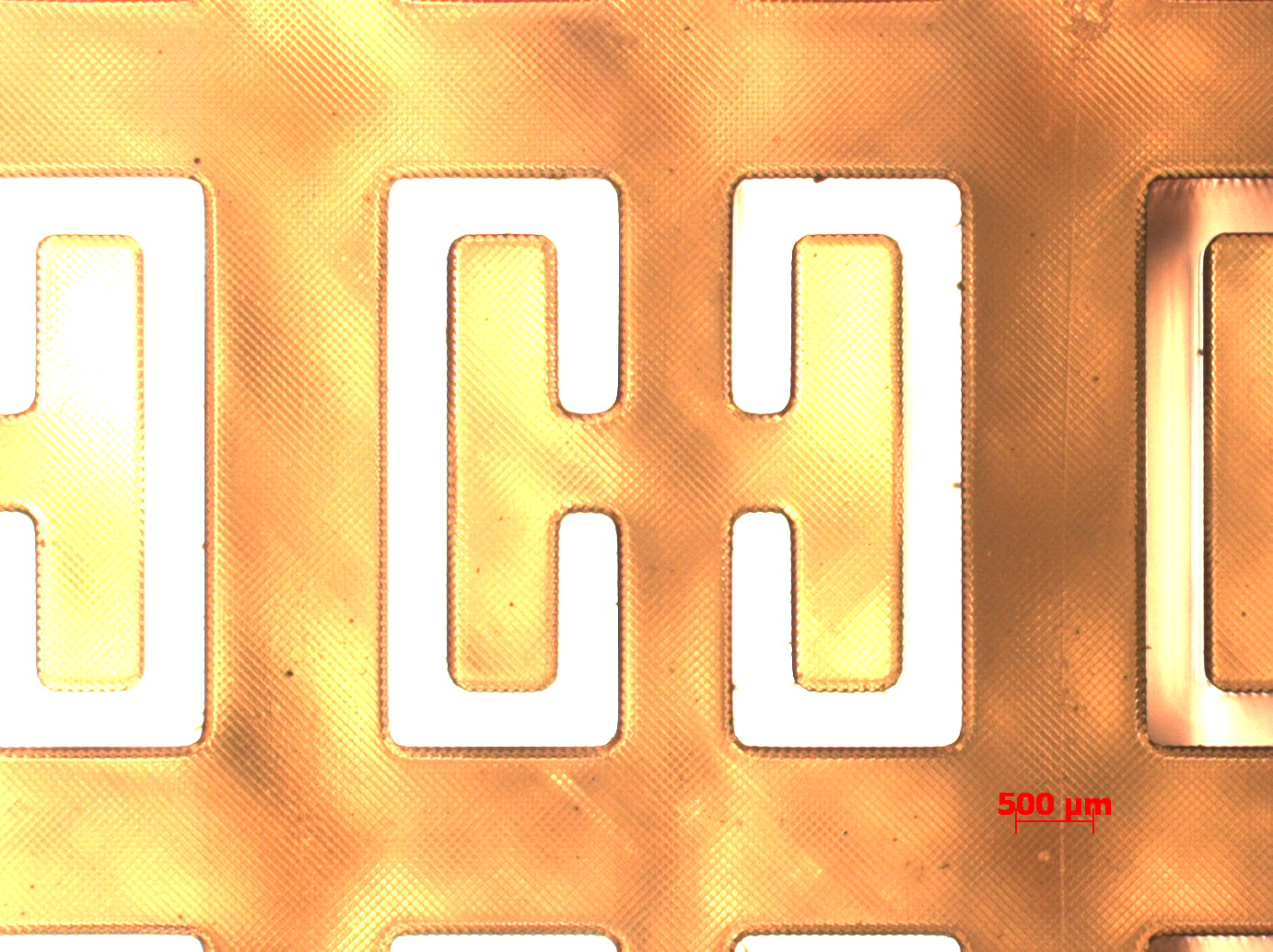}
    \caption{}
    \label{pr48_micro}
  \end{subfigure}%
  ~ 
  \begin{subfigure}[htb]{0.35\textwidth}
    \centering
    \includegraphics[scale=0.27]{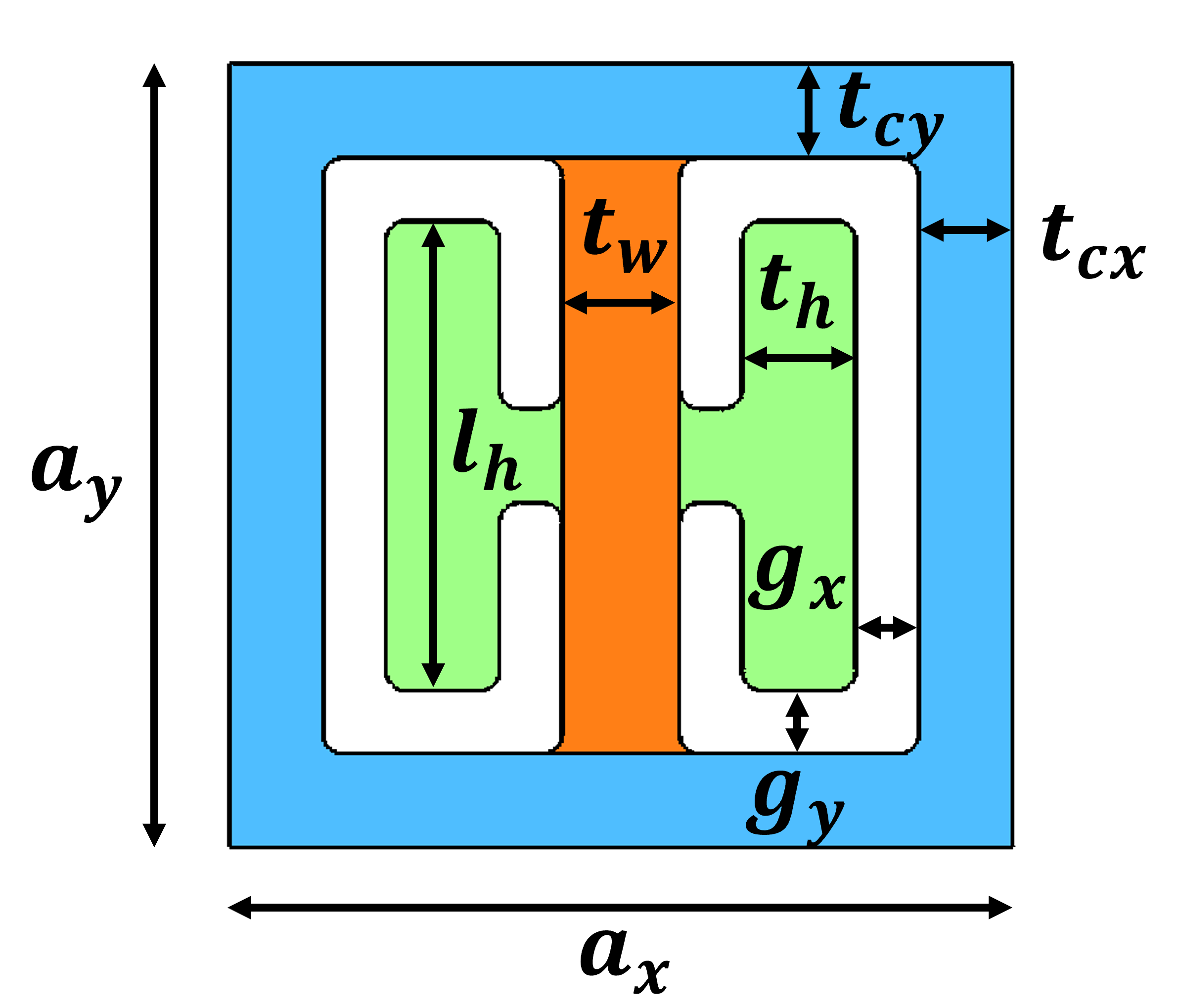}
    \caption{}
    \label{H_notation}
  \end{subfigure}
  \caption{\subref{pr48_micro}) Microscope image of an H-shaped resonator in a square cell printed with PR48 resin; see Section~\ref{sec_param_obj} for detail. \subref{H_notation}) Notation for identifying geometric features. Recurring features are only marked once to simplify the figure, e.g. $g_x$ applies to all horizontal gaps both left and right of the two head widths $t_h$.}
  \label{H_shaped_cell}
\end{figure*}

The optimization case performed in \cite{morris_expanding_2022} included X-shaped resonators in square cells with a dispersion band gap extracted from a continuum eigenfrequency model of a single unit cell with periodic boundary conditions. While useful for optimizing band width and center frequency as well as discovering design trends, such an approach is limited in practicality due to the periodically infinite domain of the eigenfrequency method. Information about travelling waves within the stop band and effects related to finite sample length typically require FEM modelling of the entire finite array for accurate predictions \cite{aghighi_low-frequency_2019,morris_multi-point_2022}. This work seeks to build upon the previous optimization approach by adding transmission calculation of a finite number of cells with more complex geometry based on ROM. The H-shaped resonator geometry shown in Figure~\ref{H_shaped_cell} is considered as a starting point, which includes the noted geometric degrees of design freedom. In particular, there is improved geometric decoupling of wall, web, and head features (blue, orange, and green domains in Figure~\ref{H_notation}, respectively) compared to the X-shaped cell. An image of the 3D printed H-shaped cell patterned with the target dimensions listed in Table~\ref{tab_pr48_H} is displayed in Figure~\ref{pr48_micro}. The average measured dimensions of the sample are also tabulated, having been extracted from optical microscope images of a 3D printed 12 cell specimen. The impact of the difference from target and standard deviation is incorporated into the model later; see Section~\ref{sec_sens}.

\begin{table}[!htb]
	\centering
	\caption{Target and average dimensions of geometric features of H-cell. Dimensions are extracted from microscope images of a PR48 3D printed specimen containing 12 H-cells.}
	\label{tab_pr48_H} 
	\begin{tabular}{cccccc} \hline
	 	                        &Target & Difference &Error &Standard\\
	 	Feature                 &Dimension &from Target &from Target &Deviation \\
	 	                        & ($\mu$m) &  ($\mu$m) &  &($\mu$m)\\ \hline
		Web Width, $t_w$  	&750    &-24    &-3.2\%		&16\\
		Head Width, $t_h$  	&700    &-18    &-2.5\%		&12\\
		Head Height, $l_h$  	&3000	&-76    &-2.5\%	    &15\\
		Gap Width, $g_x$      &400	&-21    &-5.3\%     &14\\
		Gap Width, $g_y$      &400	&-48    &-12.0\%	&16\\
		Wall Thickness, $t_{cx}$   &600	&-8     &-1.3\%		&10\\
		Wall Thickness, $t_{cy}$   &600	&2      &0.3\%		&16\\
		Cell Width, $a_x$       &5000	&-158   &-3.2\% 	&7\\
		Cell Width, $a_y$       &5000	&-168   &-3.4\% 	&27\\   \hline
	\end{tabular}
\end{table}

\subsection{Design Parameters and Objectives} \label{sec_param_obj}

\begin{figure}[!htb]
    \centering
    \includegraphics [scale=0.5]{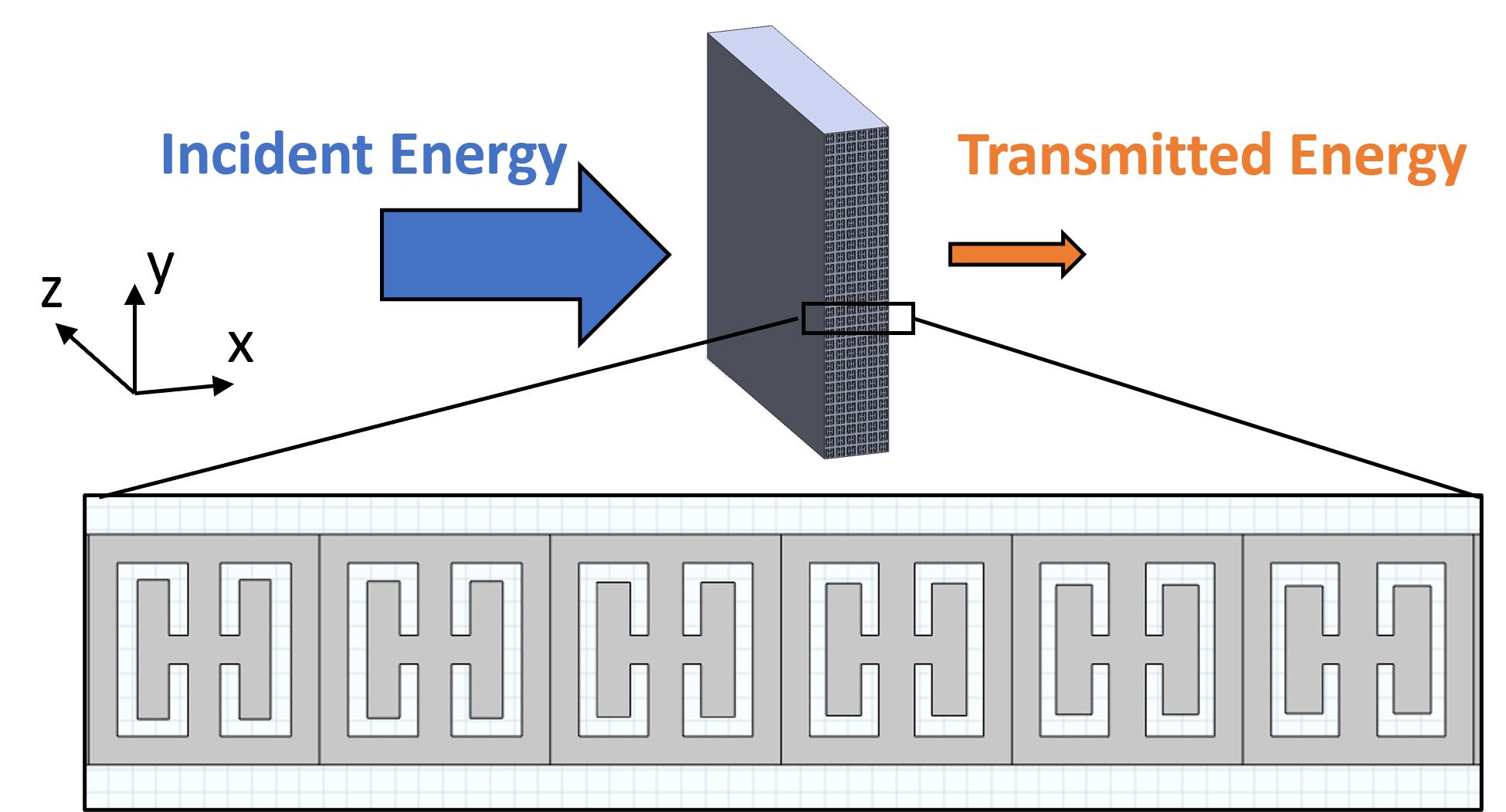}
    \caption{Finite slab of 6 cells and transmission problem. The structure is infinitely periodic in $y$ direction and in plane-strain conditions (normal to $z$ axis).}
    \label{trans_objective} 
\end{figure}

To maintain practically applicable results, the bounds of the study are constrained to proven manufacturable conditions. The constituent material properties of the metamaterial cells are for the 3D printed AutoDesk PR48 resin with Young's modulus $E=2.43$ GPa, density $\rho=1.23$ g/cm$^{-3}$, Poisson's ratio $\nu=0.35$, and loss $\eta_s=2\%$ \cite{shah_highly_2020}. The outside media, in which the metamaterial array is placed, is aluminum with $E_m=70$ GPa, $\rho_m=2.7$ g/cm$^{-3}$, and $\nu_m=0.33$. The structure is periodic out-of-plane with only one 5 mm cell modelled in $y$ and infinite length along the $z$ axis (modelled as plane strain). A slab with the 2D metamaterial microstructure as depicted in Figure~\ref{trans_objective} has 6 cells through the thickness ($x$ direction). This thickness provides an adequately thin medium (30 cm) for applications but enough cells such that graded metamaterial performance can be achieved through variation of each cell.

Following the limitations observed with printed specimens in \cite{morris_expanding_2022}, the geometric bounds are defined in Table~\ref{tab_geom_bounds} where minimum values for gaps and wall thicknesses of 300 and 250 $\mu$m, respectively, are maintained. The upper bounds (and lower for $l_h$) come from the build envelope limitations.

\begin{table}[!htb]
	\centering
	\caption{Geometric bounds for cell features, determined by minimum manufacturable size/resolution or build envelope limitations.}
	\label{tab_geom_bounds} 
	\begin{tabular}{ccc} \hline
	    Feature & Lower Bound & Upper Bound\\
	 	                        & ($\mu$m) &  ($\mu$m)\\ \hline
		Web Width, $t_w$  	&250    &1000\\
		Head Width, $t_h$  	&250    &1300\\
		Head Height, $l_h$  	&600	&3300\\
		Wall Thickness, $t_{cx}$   &250	&1000\\
		Wall Thickness, $t_{cy}$   &250	&1000\\   \hline
	\end{tabular}
\end{table}

The primary design objective is to minimize transmission within a specified target frequency band. In terms of practical application, this situation creates a wide frequency range in which the majority of the travelling wave energy does not pass through the slab and is instead reflected back to the source or dissipated within the specimen. Where trade-offs in the attenuation band width and depth are observed, a secondary objective prioritizes band width over additional depth below $10^{-6}$. This choice comes from observations during experimental characterization, where signal amplitudes cannot be resolved from noise when the transmission coefficient is lower than $10^{-6}$ \cite{morris_characterization_2023}. 
The genetic algorithm is used to optimize an objective function based on 18 geometric variables (6 cells with 3 geometric variables each: $t_w$, $t_h$, $l_h$). This choice of design space stems from the lower number of parameters and ease of presenting visual examples, i.e. resolving feasible geometries after finding GA solutions. Permitting wall thicknesses $t_{cx}$ and $t_{cy}$ to vary would require calculation of ROM parameters of a cell based on not only its own parameters but also those of neighboring cells. While this calculation is in principle possible, it adds unnecessary complexity to the present work. Finally $g_x$ and $g_y$ are calculated to ensure uniform values of $a_x$ and $a_y$ throughout the array. Solving GA in terms of ROM parameters is also useful and would require the construction of a secondary robust tool capable of translating $\beta$ and $M$ parameters into 3D printable terms, while enforcing geometric, material, or any other design criteria. For a single material print, this translation is mostly a geometric inverse design problem, which has been discussed in \cite{morris_expanding_2022}. Especially when the number of available geometric design parameters are much higher than that of the ROM parameters, this geometric translation is a desirable process, and would potentially provide a class of potential geometries for a particular ROM cell. Finally, the objective function consists of integrals of the transmission curve, sectioned and weighted over the frequency range. Three variations are used and the unique optimized arrays for each are discussed in Section~\ref{sec_results}. The solver is arranged to continue advancing generations until the mean fitness of all current children stops improving beyond a fine tolerance.

\subsection{ROM Parameters - Mass and Stiffness}

Following the methods outlined in \cite{morris_expanding_2022}, the geometric dimensions can be translated to representative ROM parameters using a series of FEM models and regression. Direct calculations of ROM parameters are made on a 3 dimensional grid of the geometric variables (9 steps for each geometric variable $t_w$, $t_h$, $l_h$). Polynomial regression fits are made to analytically relate geometry to ROM mass and stiffness parameters in the entire printable region. This regression avoids the need to run FEM during every function call while maintaining high accuracy for approximated ROM parameters. In this process, the cell response is determined based on four ROM parameters, representative of only the longitudinal vibrational mode (2 degrees of freedom), including frame mass $M_f$ and stiffness $\beta_f$ as well as resonator mass $M_r$ and stiffness $\beta_r$. The full range for each parameter is shown in Table~\ref{tab_ROM_bounds}, which bound the genetic algorithm solution for a single material system.

\begin{table}[!htb]
	\centering
	\caption{Bounds for ROM parameters determined from FEM models with the geometric parameter ranges listed in Table~\ref{tab_geom_bounds}. The values are shown assuming unit (1 m) thickness.}
	\label{tab_ROM_bounds} 
	\begin{tabular}{ccc} \hline
	    ROM Parameter                   & Lower Bound & Upper Bound\\ \hline
		Frame Mass, $M_f$ (kg) 	            &0.0135      &0.0163 \\
		Frame Stiffness, $\beta_f$ (N/m)     &$3.12\times10^8$	&$2.78\times10^9$ \\
		Resonator Mass, $M_r$ (kg)          &0.0016 	    &0.0113 \\
		Resonator Stiffness, $\beta_r$ (N/m) &$5.75\times10^6$	&$4.40\times10^8$ \\   \hline
	\end{tabular}
\end{table}

\subsection{Transmission Calculation} \label{sec_trans_calc}

A description of transmission $|\mathsf{S}_{ab}{}^2|$ as a function of longitudinal ROM parameters $M_f$, $\beta_f$, $M_r$, and $\beta_r$ is discussed in \cite{wang_exceptional_2022} with the relevant equations transcribed here. A transfer matrix is constructed individually for each of the 6 cells ($n=1,2,...6$) using
\begin{equation}
    \mathsf{T}=\begin{bmatrix} 1-\frac{K_\mathsf{T}}{2\beta_f} & \im\omega(\frac{4\beta_f-K_\mathsf{T}}{4\beta_f^2}) \\ \frac{\im K_\mathsf{T}}{\omega} & 1-\frac{K_\mathsf{T}}{2\beta_f} \end{bmatrix},
\end{equation}
where $\omega = 2 \pi f$ is the angular frequency of the incident harmonic wave and
\begin{equation}
    K_\mathsf{T}=\frac{M_r\omega^2}{1-(\omega/\omega_r)^2}+M_f\omega^2,
\end{equation}
and the natural resonance of the resonator is given by 
\begin{equation}
    \omega_r=\sqrt{\beta_r/M_r}.
\end{equation}
For a wave travelling through the slab left to right along $x$ axis, the leftmost cell is numbered $n=1$ and the rightmost is $n=6$. The transfer matrix for the full array ($\mathsf{TM}$) is then assembled from the individual contributions of each cell
\begin{equation}
    \mathsf{TM}=\mathsf{T}_6 \mathsf{T}_5 \mathsf{T}_4 \mathsf{T}_3 \mathsf{T}_2 \mathsf{T}_1.
\end{equation}
To calculate the transmission coefficient of this discrete system, the elastic wave impedance of the outside media must be multiplied by the cross section area of the unit cell $A_m$:
\begin{equation}
    Z_m=A_m\sqrt{E_m \rho_m}.
\end{equation}
The effective modulus, $E_m$, depends not only on the bar material, but also on the desired boundary conditions of the model. For a bar with realistic proportions $A_m$ and for the results of this paper, the assumption that $E_m$ is simply the Young's Modulus is satisfactory. Ultimately, the effect on design optimization is also minimal as the same value is used for all cases, and that the depth of transmission curves is dominated by the resonance features of the cells. The scattering coefficient $\mathsf{S}$ can then be determined, specifically
\begin{equation} \label{eq:sab}
    \mathsf{S}_{ab}=\frac{2Z_m}{\mathsf{TM}_{21}+\mathsf{TM}_{11}Z_m+\mathsf{TM}_{22}Z_m+\mathsf{TM}_{12}Z_m^2},
\end{equation}
and the power transmission amplitude will be simply $T = |\mathsf{S}_{ab}{}^2|$.

\begin{figure}[!htb]
    \centering
    \includegraphics [scale=0.3]{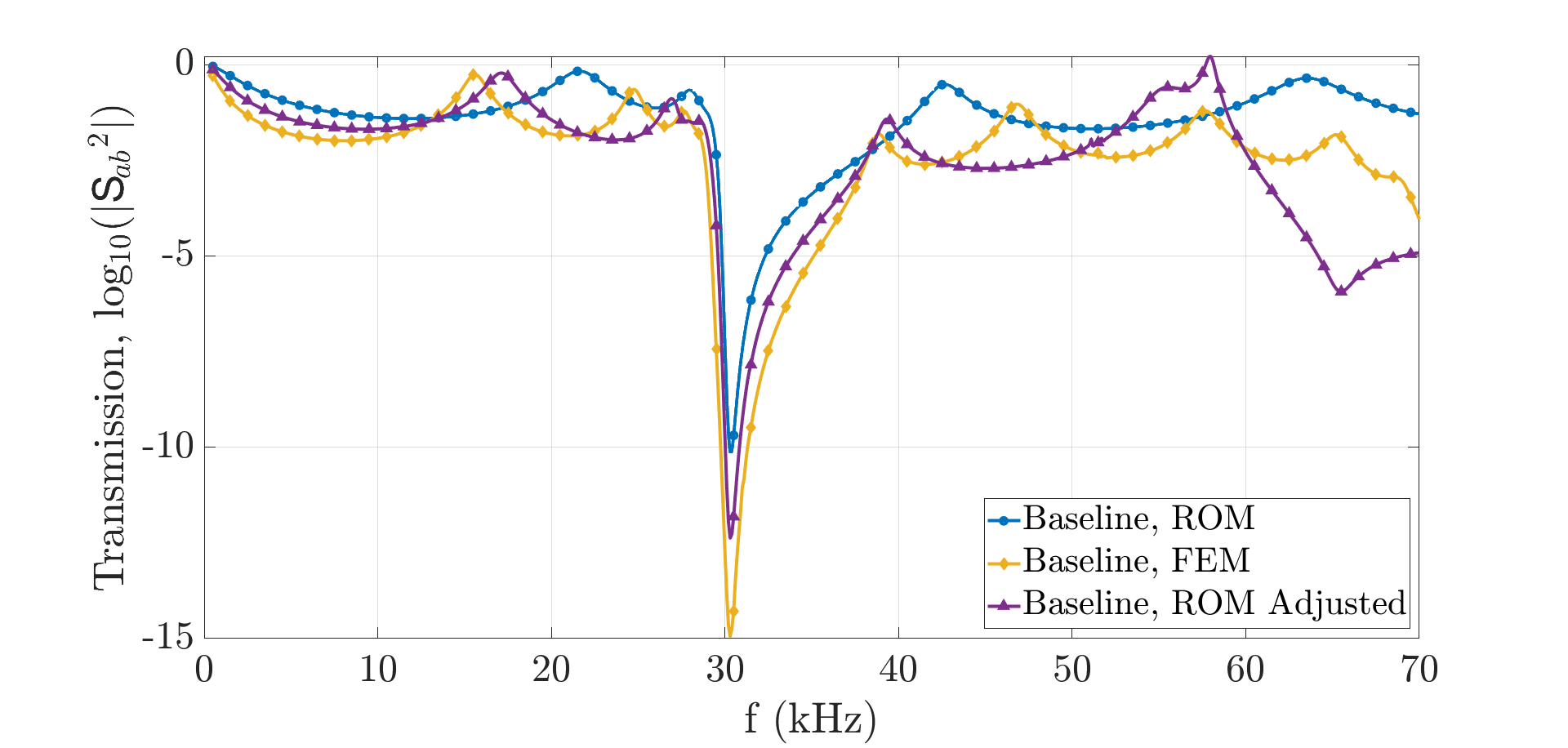}
    \caption{Validation of the 2 DOF ROM approximation using a corresponding FEM solution from a 1D scattering model, as well as a version of the ROM with edge cell adjustments as described in Section~\ref{sec_trans_calc}.}
    \label{baseline_trans} 
\end{figure}

\begin{figure}[!htb]
    \centering
    \includegraphics [scale=0.5]{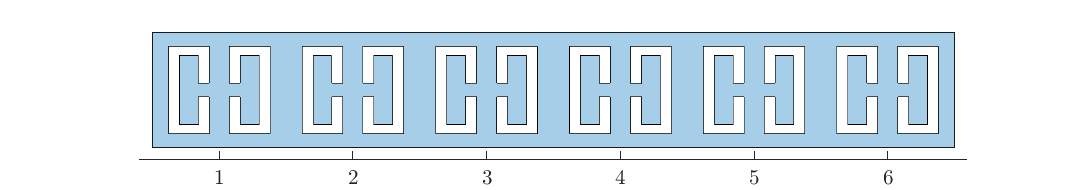}
    \caption{The baseline array constructed of 6 identical H-shaped cells.}
    \label{baseline_cells} 
\end{figure}

For the baseline array of 6 H-shaped cells with the dimensions listed in Figure~\ref{tab_pr48_H}, the transmission curve was approximated with ROM and compared with FEM to validate the accuracy of the tool. The overlay of the ROM with FEM solutions is provided in Figure~\ref{baseline_trans} along with a visualization of the array with 6 identical baseline cells in Figure~\ref{baseline_cells}. The FEM result follows the method outlined in \cite{morris_multi-point_2022} for the 1D scattering of an ultrasonic cell, with the aluminum outside media properties and baseline cell dimensions from this work substituted in. For perspective, the FEM result required about 1 hour to generate while the ROM calculates its full curve in under 1 second. The simplification of the problem that leads to orders of magnitude time savings also develops some differences in the solutions. The locations and amplitude of peaks outside of the stop band (0 to 28 and 42 to 70 kHz) vary. The response outside of the stop band is driven by DOFs that are ignored by ROM and are not of great interest for optimization of a longitudinal stop band. Specifically, these DOFs are prominent at the edge cells (1 and 6). Within and around the stop band, the center frequency and lower band limit are accurately predicted with a small variation in the upper band limit. This shift in the upper limit, and the difference in depth of the two valleys, are related to the treatment of edge cells as well. The cells in positions 1 and 6 (outside edges of the array) have different effective ROM parameters and impedance than their interior counterparts. FEM includes this impedance difference but ROM does not, due to the boundary with uniform external medium, thus creating deviation in their transmission amplitudes. This effect can be verified by exchanging the transfer matrices of only the outer cells in the ROM model with their FEM equivalents, found following the 2 cell boundary cell procedure in \cite{morris_multi-point_2022}. The transfer matrix of the full 6 cell array is then recalculated, consisting of 2 outer cells from FEM and 4 interior cells from ROM, rewritten as 
\begin{equation}
    \mathsf{TM}'=\mathsf{T}'_6 \mathsf{T}_5 \mathsf{T}_4 \mathsf{T}_3 \mathsf{T}_2 \mathsf{T}'_1
\end{equation}
with the transfer matrices taken from FEM denoted as $\mathsf{T}'$. When applied and presented in results, this technique will be identified to as 'ROM Adjusted'. The transmission curve can then be calculated as usual using Eq.~\ref{eq:sab} and is plotted in Figure~\ref{baseline_trans}. 
When adjusting for the boundary cells' effects, the upper band limit now matches exactly with FEM and the depth of the transmission curve is much closer to the FEM prediction. The low frequency range (up to 30 kHz) matches closely for FEM and ROM Adjusted; the higher frequency range (greater than 40 kHz), however, has increased variation. The ROM is expected to predict band location and width well enough such that the time savings justify the high frequency mismatched aspects. Since the edge cell effect is consistent for all ROM solutions, design trends and paths towards optimized performance are also expected to be meaningful. Therefore, in the following, the optimization of the array is performed with a purely ROM-based approach, with the final selected cases being reanalyzed to include and evaluate the effects of the boundary or edge cells properly. In the future, an adjustment of the ROM parameter for these cells is planned so that the final performance metrics curves match the full field results (FEM) as closely as possible. However, this improved accuracy will require an extended computation time. Ultimately, if the interior cells' transfer matrix were also calculated according to \cite{morris_multi-point_2022}, the match between FEM and transfer matrix analysis would have been nearly perfect as discussed in that reference.

\section{Results and Discussion} \label{sec_results}

\subsection{Optimization for a Deeper Stop Band}

\begin{figure}[!htb]
    \centering
    \includegraphics [scale=0.3]{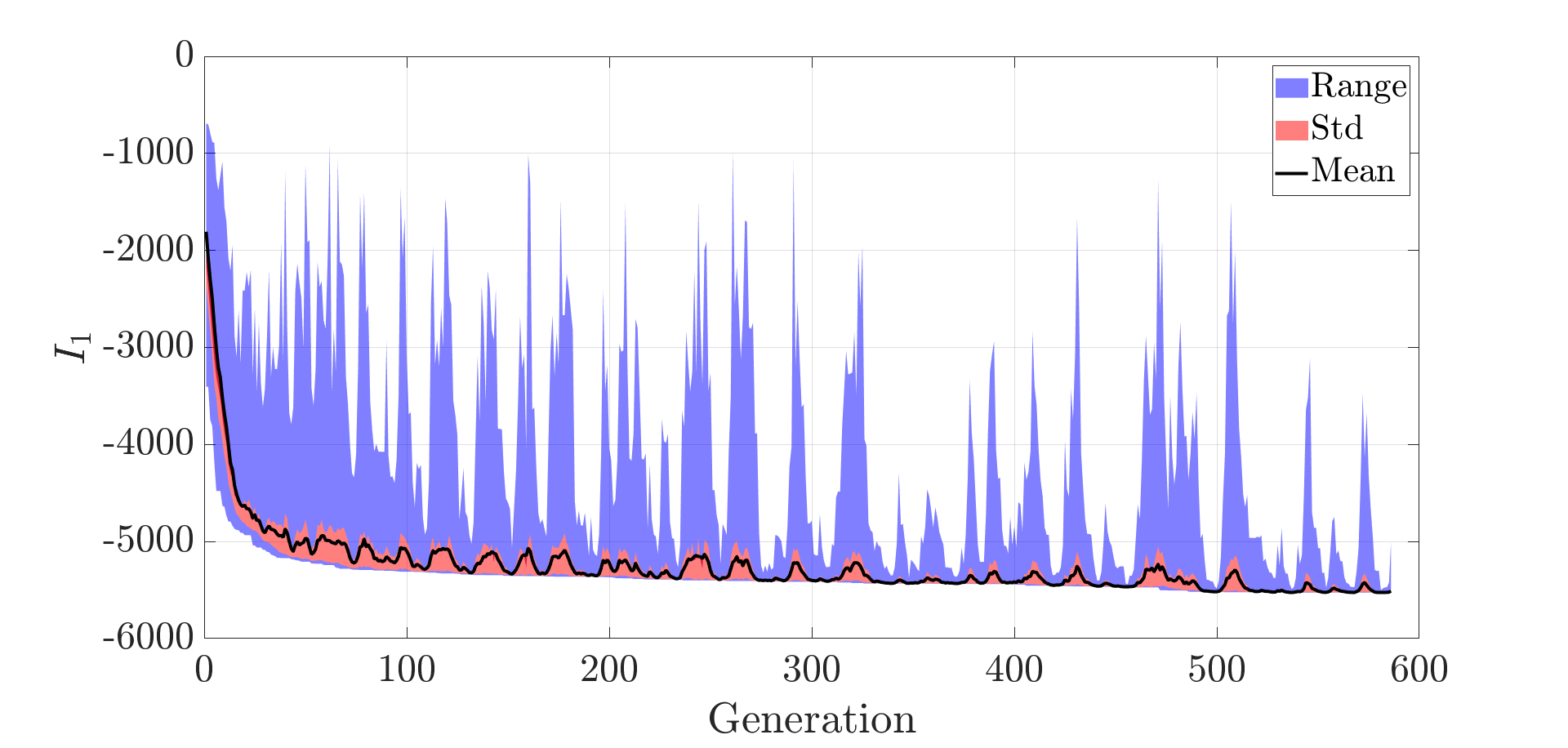}
    \caption{The GA converges to a solution in 585 generations. Exploitation and exploration can be visually approximated as standard deviation and range increases, respectively.}
    \label{conv_shaded} 
\end{figure}

Starting from the baseline array shown in Figure~\ref{baseline_cells}, optimization of the array's transmission performance was conducted. To assess manufacturing feasibility directly and for better visual comprehension, the optimized designs for the 6 cells were generated using 18 geometric parameters ($t_w$, $t_h$, and $l_h$ for each cell). As a theoretical exercise, the ROM parameters could be used as genes, but a robust inverse solver that translates ROM parameters to feasible geometries would need to be developed (outside the scope and generality desired in the present work). Model performance is measured in terms of applicability, where the stop band targets for depth and width have been met via feasibly manufacturable cell geometries. The first optimization scheme attempts to attenuate transmission energy over the same approximate frequency range as the baseline design, by integrating log-scale transmission between 30 and 40 kHz. The initial performance metric formula is written as
\begin{equation} \label{eq:3040mT}
I_1 = \int_{30~\text{kHz}}^{40~\text{kHz}} \log_{10} |\mathsf{S}_{ab}{}^2|~df.
\end{equation}
To solve this problem, GA required 111360 function calls over 585 generations until the solver was no longer measuring improvement of a tolerance greater than $10^{-6}$ (10 minutes computation time using 64 cores). The convergence of the function is plotted in Figure \ref{conv_shaded}, where the exploitation (children based on best performers) and exploration (random mutations testing unique gene combinations) can be seen as standard deviation and range increases, respectively. The standard deviation as shown is determined as the closest 68.2\% of the children to the mean.

\begin{figure}[!htb]
    \centering
    \includegraphics [scale=0.3]{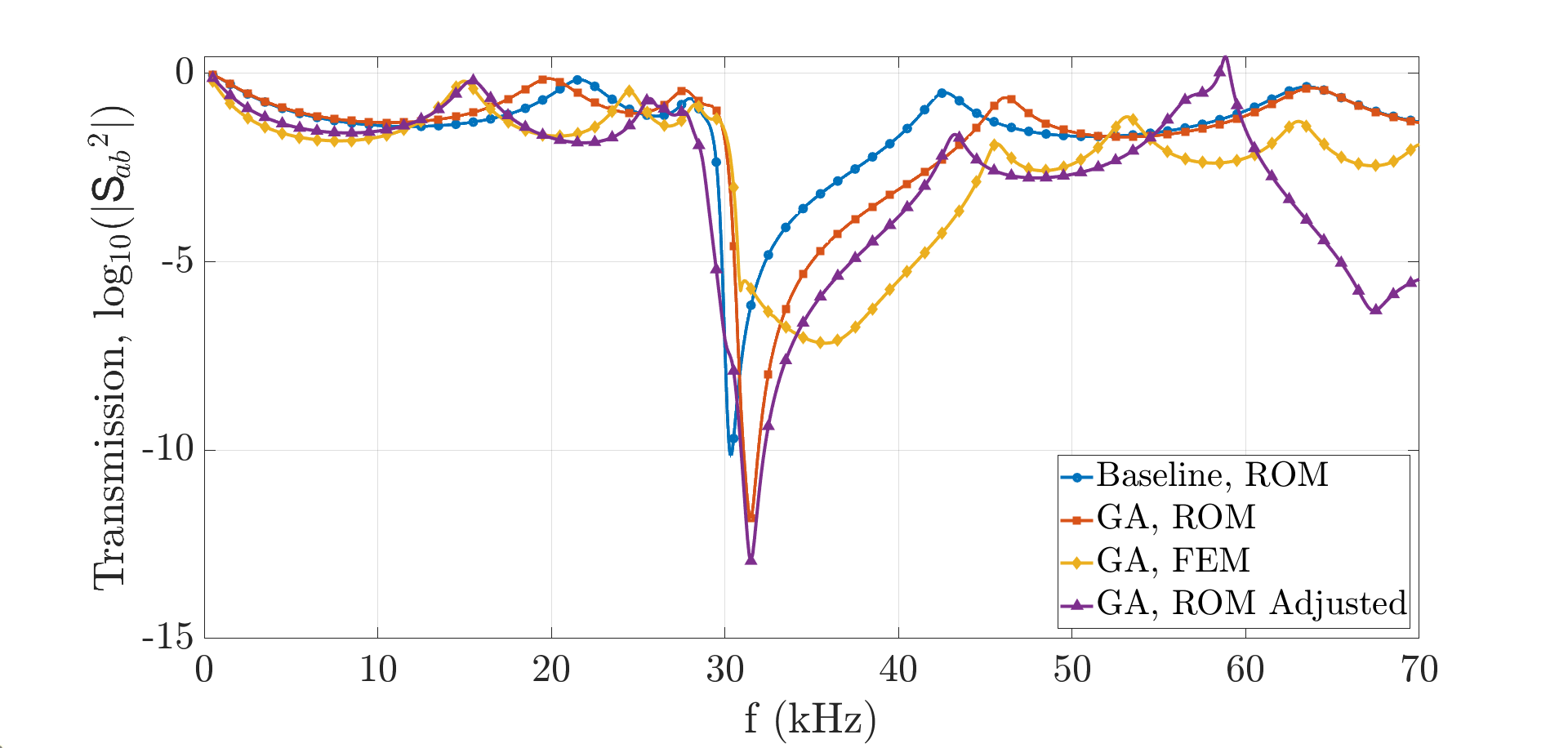}
    \caption{Transmission curves for the GA solution minimizing the objective function defined in Eq.~\ref{eq:3040mT}.}
    \label{3040_trans} 
\end{figure}

The ROM-based transmission curve for the best specimen from GA is shown in Figure~\ref{3040_trans}, alongside the ROM-based baseline result, where two orders of magnitude improvement in depth are observed. The FEM results for the GA-optimized array were computed for verification and highlights some qualities of the ROM method. Again here the bounds of the stop band match well, however the depth of the curve varies more dramatically than was observed previously in Figure~\ref{baseline_trans}. With adjustment of the edge cells' transfer matrix, the transmission curve of the GA-optimized array (also shown) provides further apparent benefit as it pushes the depth of the curve towards the desired solution in the 35-40 kHz region but does not correct the overall shape mismatch of the curve. This highlights an overall disagreement between ROM and FEM transfer matrix predictions that warrants further investigation in a separate study focusing on improving the performance of ROM both at interior as well as the edge cells.

\begin{figure}[!htb]
    \centering
    \includegraphics [scale=0.5]{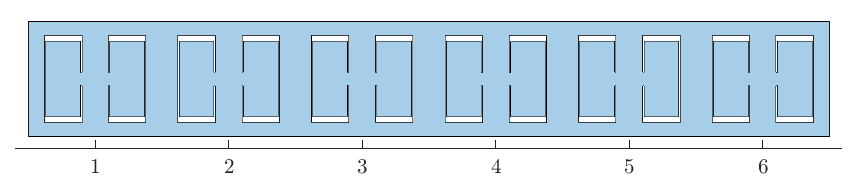}
    \caption{Reconstructed array based on the GA optimized ROM parameters for the objective in Eq.~\ref{eq:3040mT}.}
    \label{3040_cells} 
\end{figure}

The recreation of the array geometry (Figure~\ref{3040_cells}) initially suggests that all cells are nearly identical with feature sizes equal to the upper geometric limits (see Table~\ref{tab_geom_bounds}); however, there are small 5-20 $\mu$m variations in a few of the features. The two outer cells in positions 1 and 6 have slightly narrower web widths of about 981 and 983 $\mu$m, respectively, compared to cells 2 to 5 which are at the upper limit of 1000 $\mu$m. Two center cells in positions 3 and 4 have almost negligibly thinner head widths of 1295 and 1297 $\mu$m, respectively, compared to the other cells (1, 2, 5, and 6) which are at the upper limit of 1300 $\mu$m. Note that these variations are considered to be functionally identical and within the printer's feature tolerance, discussed in more depth in Section~\ref{sec_sens}.

\begin{figure}[!htb]
    \centering
    \includegraphics [width=\textwidth]{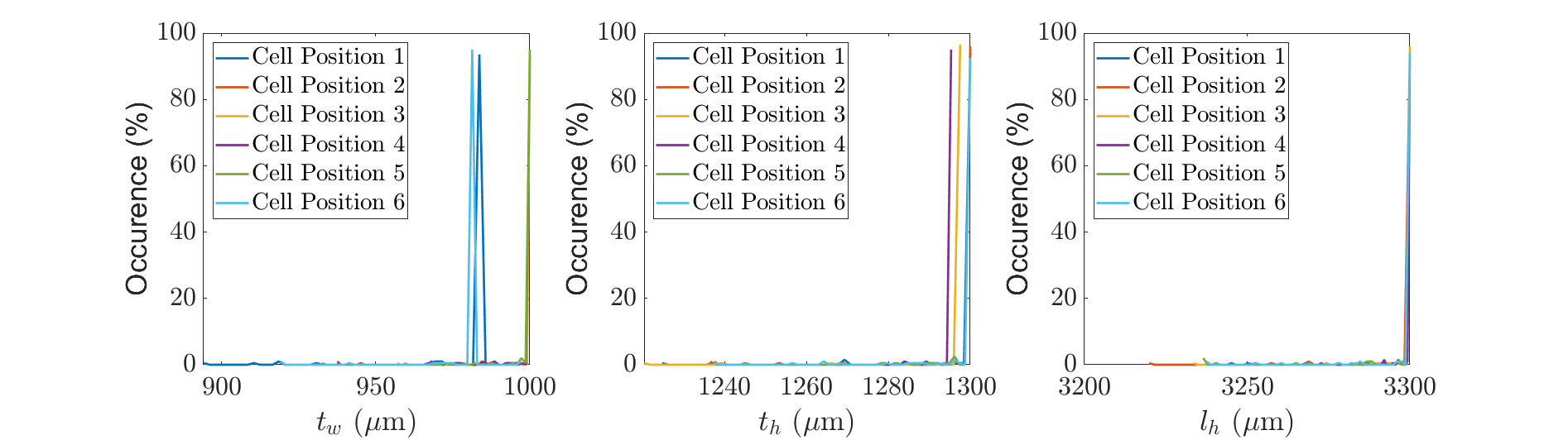}
    \caption{Features from the genome structure of the final GA generation yield a converged solution with cell position dependent qualities. $N=200$}
    \label{genome} 
\end{figure}

The geometric differences of the array in Figure~\ref{3040_cells} are more apparent when looking at the genome structure of the final generation for the GA solution, provided in Figure~\ref{genome} . Each sub-figure depicts the rate of occurrence of each trait for the final 200 arrays, where the majority of the values center around a converged solution. The small variations in head width are now clear, with the center cells (positions 3 and 4) located within 5 $\mu$m below the other positions which had converged to 1300 $\mu$m. The web width for the two outer cells (positions 1 and 6) converges to around 980 $\mu$m, different from the 1000 $\mu$m that the four interior cells all converge toward. In \cite{morris_multi-point_2022}, the boundary cells were observed to have response that was a function of the ambient media's properties as well as their own microstructure, while the interior cells were not affected by the ambient properties. The observed differences in the GA-optimized structures are also driven by the fact that the effect of boundary cells on the transmission is different than that of the interior cells, and further underlines that the proper optimization of the array structure requires knowledge of the specific relationship between the boundary cell and the ambient media, likely requiring careful additional adjustment of the boundary cells to compensate. Note that all arrays identified independently by GA have varied properties at their boundary cells. Refinement and careful application of these boundary effects within ROM will be subject of future work, however it is interesting that the model captures aspects of this naturally. Another important observation about the genome structure is that it has converged to a single solution. For each of the features in the example provided in Figure~\ref{genome}, the solution is centered around a single size with very few points remaining as outliers. The converged data set inspires confidence that the GA solution is not stuck in local minima. Good convergence is also the case for each of the subsequent models, so their final genome structures will not be plotted to avoid redundancy.

\subsection{Optimization for a Wider Stop Band}

\begin{figure}[!htb]
    \centering
    \includegraphics [scale=0.3]{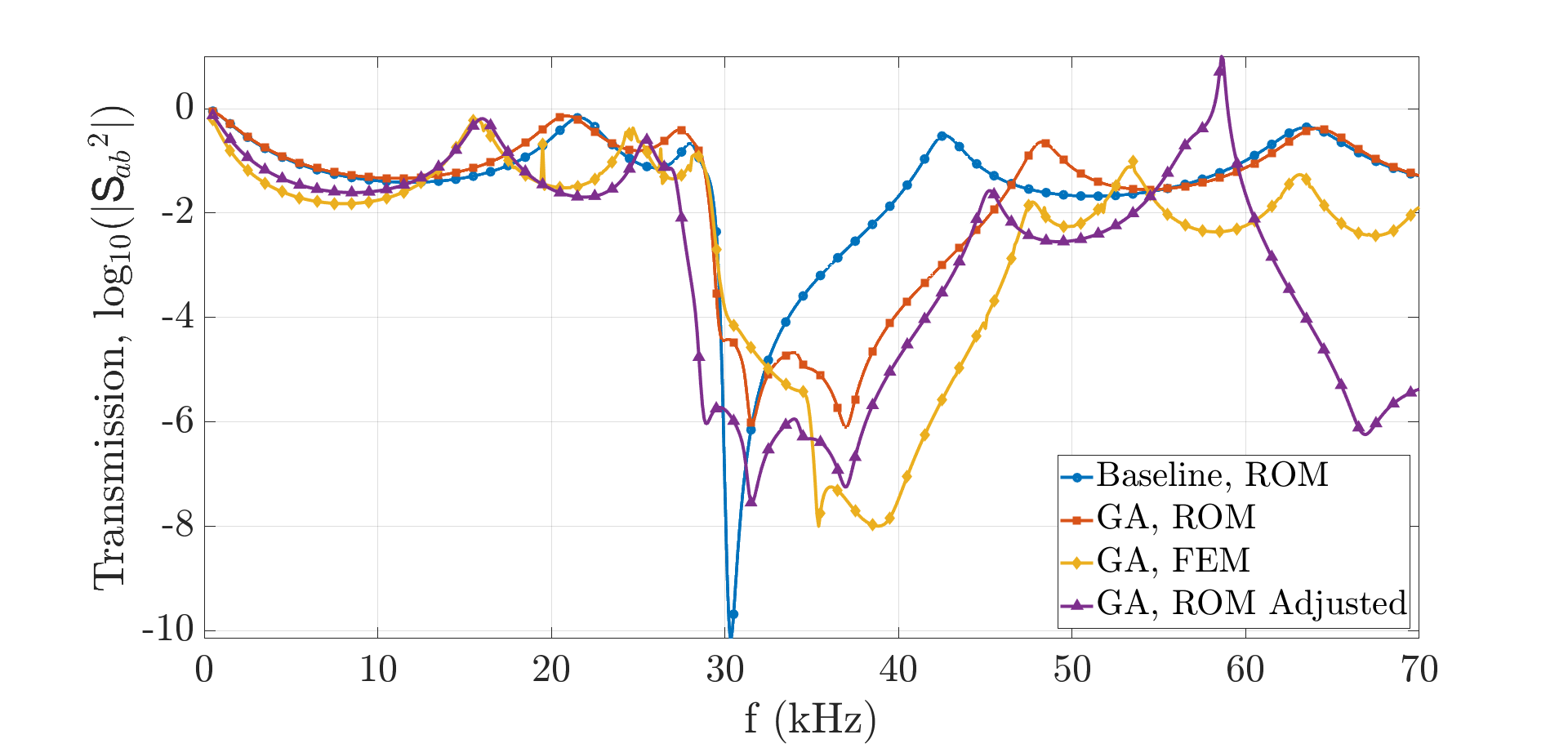}
    \caption{Transmission curves for the GA solution minimizing the objective function defined in Eq.~\ref{eq_obj1}.}
    \label{wider2050_trans} 
\end{figure}

\begin{figure}[!htb]
    \centering
    \includegraphics [scale=0.5]{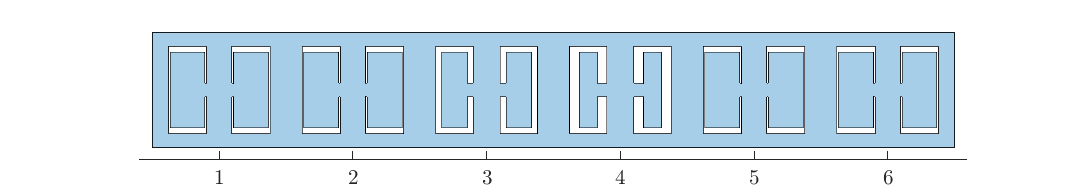}
    \caption{Reconstructed array based on the GA optimized ROM parameters for the objective in Eq.~\ref{eq_obj1}.}
    \label{wider2050_cells} 
\end{figure}

The next optimization goal example is to improve attenuation over a wider stop band, compared to the baseline array. The integration limits were set to attempt to double the band width up to a range of 30 to 50 kHz. Continuing to ignore the effects outside of the stop band for now, the objective function is defined as
\begin{equation} \label{eq_obj1}
I_2 = \int_{30~\text{kHz}}^{50~\text{kHz}} M(f)~df
\quad\text{where}\quad
M(f) = \begin{cases}
\log_{10} |\mathsf{S}_{ab}{}^2| & \log_{10} |\mathsf{S}_{ab}{}^2|\geq -6 \\
-6 & \log_{10} |\mathsf{S}_{ab}{}^2|< -6
\end{cases},
\end{equation}
where the piece-wise function serves to prioritize width over additional depth below the floor of $|\mathsf{S}_{ab}{}^2|=10^{-6}$. The genetic algorithm required 65950 function calls over 346 generations to find a solution (6 minutes computation time using 64 cores). The optimized transmission curve is plotted in Figure~\ref{wider2050_trans} along with the baseline and FEM solution and edge cell corrected version of ROM for verification. The result is successfully wider than the baseline, extending the upper bound to 48 kHz. The FEM result validates the ROM solution, matching the bounds of the stop band accurately while exceeding the predicted depth. Again here, the adjusted ROM solution with edge cell transfer matrices exchanged with scattering model solutions provides some depth corrections, but the existing ROM curve shape is pushed outward and mismatches in shape remain between ROM and FEM stop bands. The optimized geometric parameters are used to regenerate the array of 6 cells, shown in Figure~\ref{wider2050_cells}. The array has an interesting gradient and symmetry, where cells 1,2,5,6 are nearly identical with features at the upper geometric bounds. Cells 3,4 are each unique with smaller head features than the outer cells. Note that while the full array is not perfectly symmetric, if the order of the cells is flipped the transmission result is exactly the same. The GA algorithm appears to bifurcate to this particular arrangement numerically.  

\subsection{Optimization for a Square-Shaped Stop Band}

\begin{figure}[!htb]
    \centering
    \includegraphics [scale=0.3]{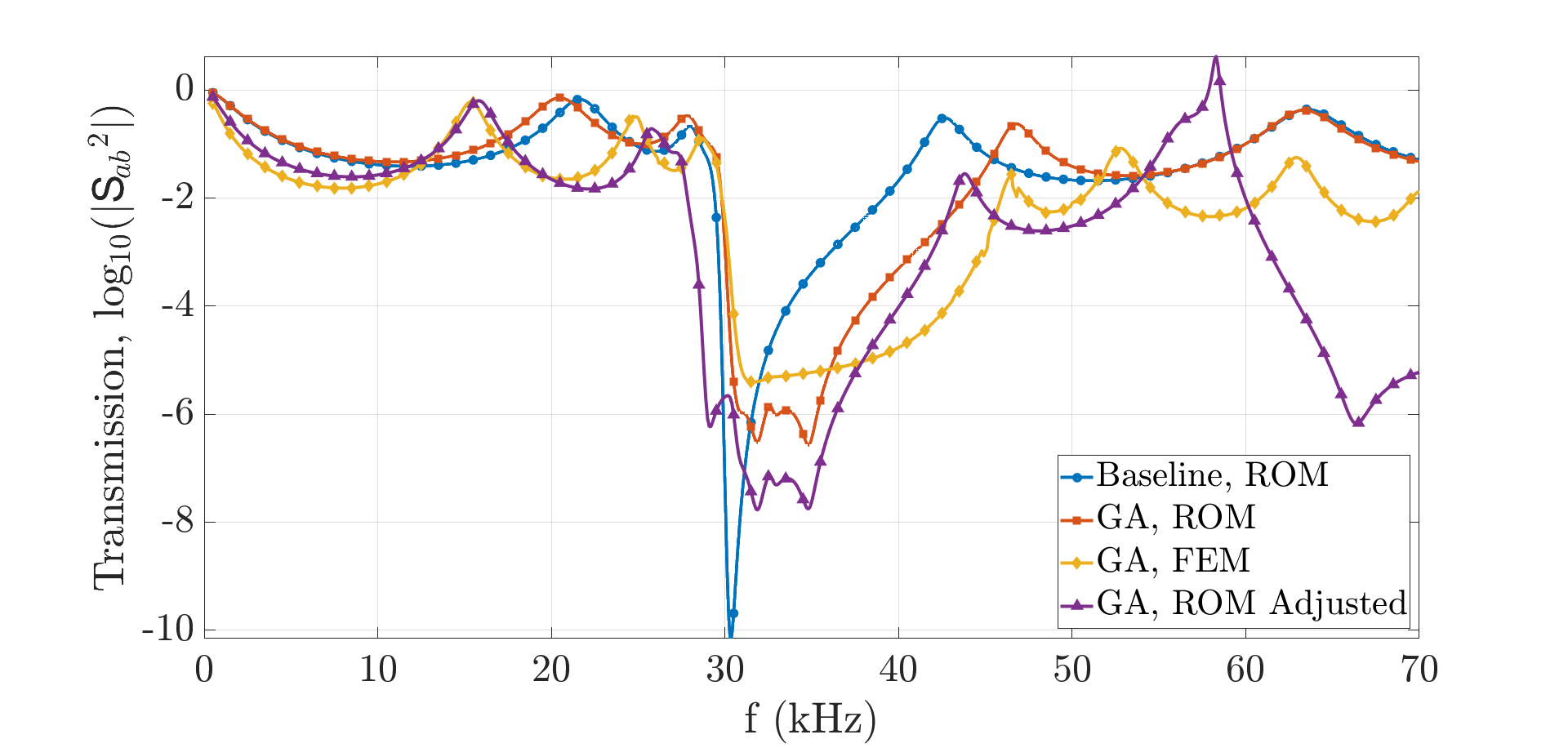}
    \caption{Transmission curves for the GA solution minimizing the objective function defined in Eq.~\ref{eq_obj2}.}
    \label{square_trans} 
\end{figure}

\begin{figure}[!htb]
    \centering
    \includegraphics [scale=0.5]{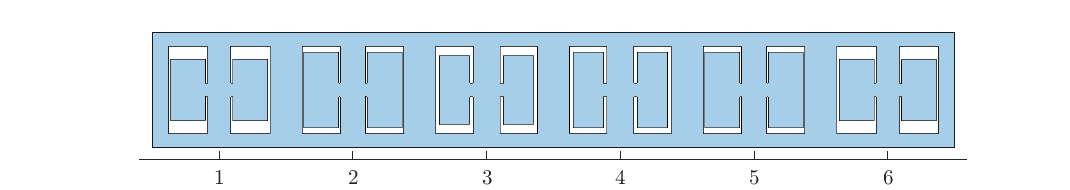}
    \caption{Reconstructed array based on the GA optimized ROM parameters for the objective in Eq.~\ref{eq_obj2}.}
    \label{square_cells}
\end{figure}

A third objective function is applied that takes into consideration the response outside of the stop band. The same piece-wise function from Eq.~\ref{eq_obj1} is split into a series of three integrals. One integral is associated with maximizing the insertion loss between 30 to 40 kHz, while the two outside of that frequency range are associated with maximizing transmission. This objective function is expected to generate a stop band with a more square shape (sharper slopes at the bounds) compared to the tapered shape of the baseline.  To do this, the two outer integrals need to have their sign changed so that the objective can minimize their values. Each integral has different length, so they are normalized by their frequency range $\Delta f$. Additionally, the two outer integrals are weighted by a factor of 2. This normalization is certainly ad-hoc and is only used for the sake of demonstration. The configuration described is written as

\begin{equation} \label{eq_obj2}
I_3 = -\frac{2}{\Delta f}\int_{0~\text{kHz}}^{30~\text{kHz}} M(f)~df
+\frac{1}{\Delta f}\int_{30~\text{kHz}}^{40~\text{kHz}} M(f)~df
-\frac{2}{\Delta f}\int_{40~\text{kHz}}^{70~\text{kHz}} M(f)~df.
\end{equation}
A solution was found after 342210 function calls and 1800 generations (30 minutes computation time using 64 cores). Success in achieving the desired square shape is seen by comparing the optimized transmission curve in Figure~\ref{square_trans} to the baseline, FEM verified solutions, and adjusted transfer matrices for edge cells. Depth between 30-40 kHz was improved compared to the baseline, although GA achieves this by extending the width beyond the 40 kHz target (up to 46 kHz). While the ROM solution still shows a tapered slope along the stop band upper bound, the trend was properly identified and the FEM solution predicts a more square shaped transmission valley with sharper bounds for this functionally graded array. Stronger weights for the regions just outside of the desired stop bands and utilizing the accurate transfer matrix for the edge cells are expected to improve the performance of this search and optimization. The reconstructed cell geometry is shown in Figure~\ref{square_cells}. Cells 1,6 and 2,5 are closely matching pairs, while 3,4 remain unique. A design trend is evident here, where an almost perfectly symmetric functionally graded structure is desired to achieve good transmission outside of a stop band with sharp bounds. Similarly to before and due to reciprocity in elastodynamics, if the cell order is reversed the transmission results remain the same.

\subsection{Sensitivity to Geometric Error or Uncertainty} \label{sec_sens}

The solutions suggested by genetic algorithm are only confined by bounds and not representative of realistic 3D printing tolerances. The printed specimen in Figure~\ref{pr48_micro} was fabricated on a DLP printer with an effective pixel resolution of 50 $\mu$m and exhibited the fabrication tolerances tabulated in Table~\ref{tab_pr48_H}. Two types of deviations from ideal geometry are present in the fabricated geometry, a uniform resin shrinkage (Mean Difference from Target) and feature repeatability (Standard Deviation). The shrinkage difference is assumed to be compensated by adjusting the dimensions sent to the printer as discussed in \cite{morris_expanding_2022}, however, the referenced analysis only considers systematic error applied to all cells independently for each feature. 

\begin{figure}[!htb]
    \centering
    \includegraphics [width=\textwidth]{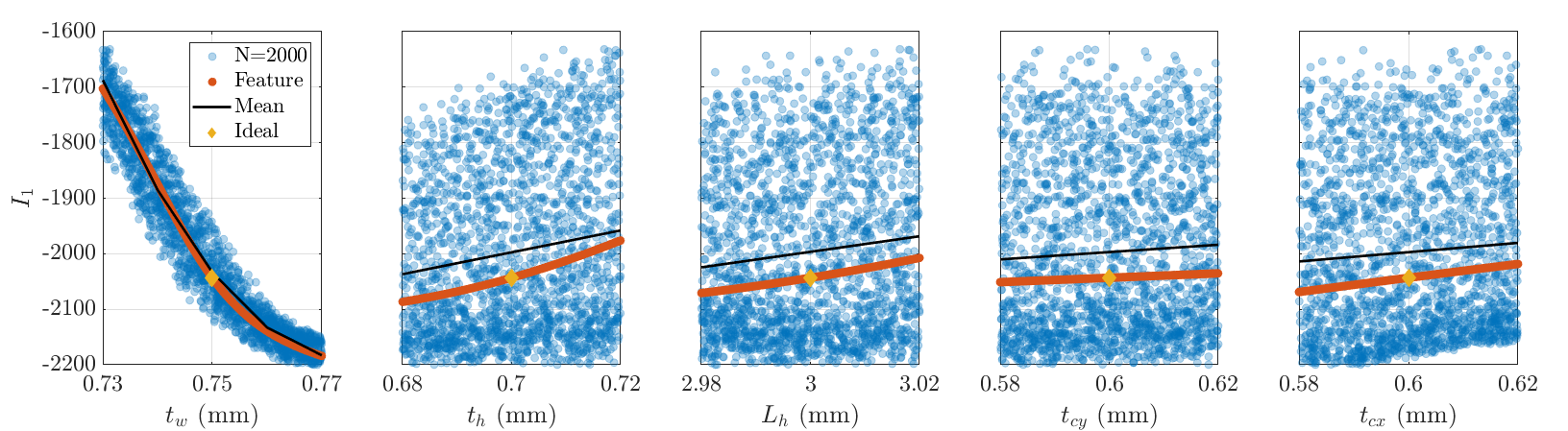}
    \caption{A random error of $\pm 20~\mu$m was introduced to each feature within 1 cell, repeated 6 times to form an array. All performance sensitivity data (blue circle) is digested using the independent feature influence (red square), full data mean (black line), and ideal array with no error (yellow diamond). $N=2000.$}
    \label{identical_performance} 
\end{figure}

The speed of the ROM enables a Monte Carlo simulation capturing the effect of repeatability error on the baseline array's transmission performance. For a standard deviation rounded to $\pm 20~\mu$m for all features, random geometric error was introduced to the web, head, and wall features of a single cell (which was repeated 6 times to form a full array). This approach allows observation of the sensitivity of the performance metric (Eq.~\ref{eq:3040mT}) as a function of printing uncertainties. The correlation plots for a population of $N=2000$ randomized arrays are presented in Figure~\ref{identical_performance}. Each blue dot represents a single specimen, the influence of the independent feature is shown by red squares (i.e. only the feature on the subplot axis has error, the other 4 features have no error), the mean of the blue data is shown as the black solid line, and the ideal array (no error) is shown as a yellow diamond. The performance is particularly sensitive to errors and uncertainties in the web width,$t_w$. The dominance of $t_w$ is apparent when comparing the lack of a gap between the mean (black lines) and feature-only variation (red lines) which is present for all other features. In contrast to $t_w$, the array is relatively insensitive to wall and head features.

\begin{figure}[!htb]
    \centering
    \includegraphics [scale=0.3]{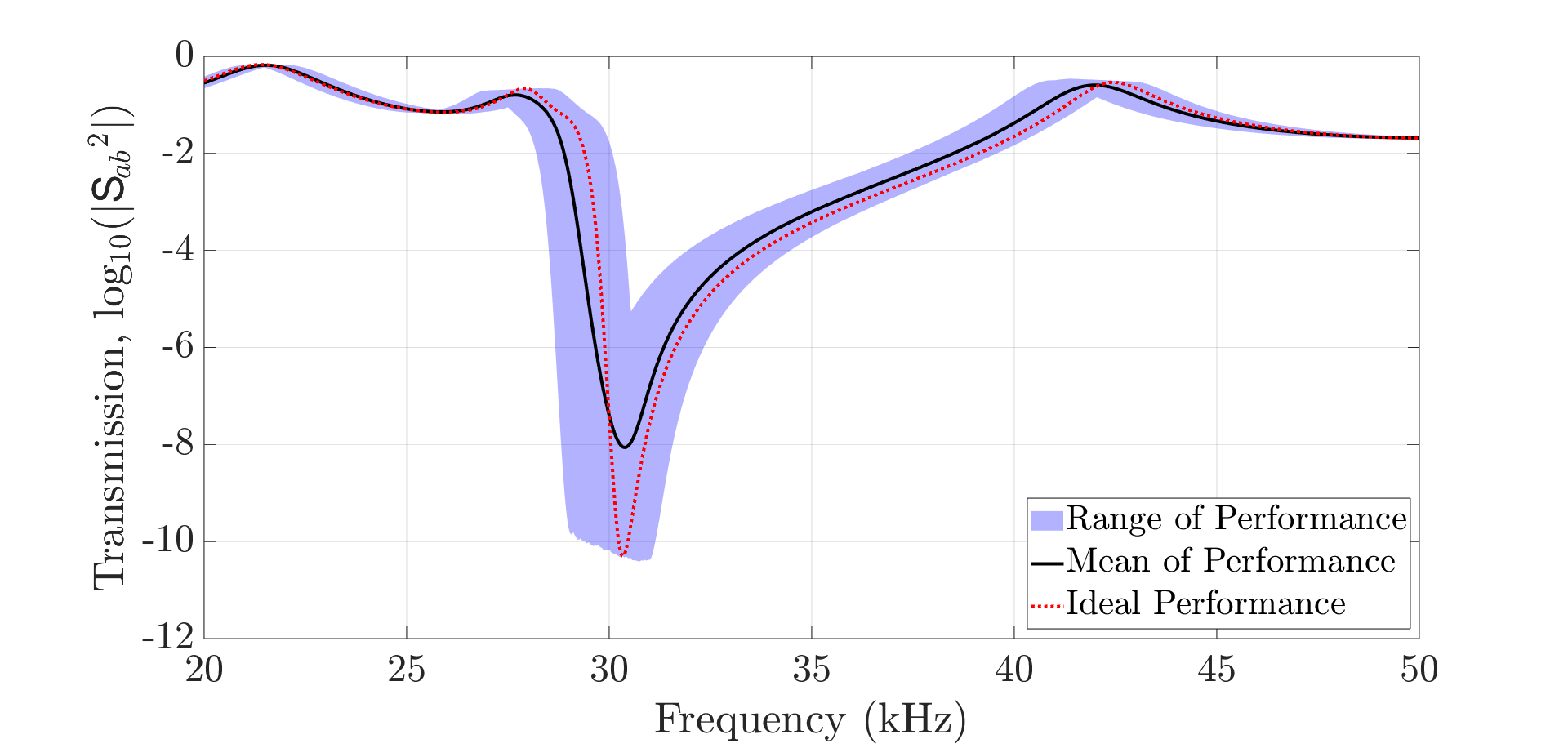}
    \caption{6 cell arrays with unique error (30 independent features) having both uniform (systematic) and randomized (uncertainty) printing errors (standard deviation of $\pm 20~\mu$m). The blue region represents every solution, the black solid line is their mean, and the red dashed line is the ideal array (no error). $N=2000.$}
    \label{measured_sensitivity} 
\end{figure}

Adding in the uniform error (Difference from Target in Table~\ref{tab_pr48_H}) helps quantify the true deviation from the ideal array. Arrays with features including both uniform error and $\pm 20~\mu$m random error were generated. In this case, each of the 6 cells has unique error (30 randomized features total) to better represent the physical specimen. The transmission curves for a randomized population of $N=2000$ arrays are overlaid in Figure~\ref{measured_sensitivity}, where the blue region includes every solution, the black solid line is their mean, and the red dashed line is the ideal array (no error). The uniform error causes a 1 kHz reduction in center frequency compared to the ideal array, along with a performance decrease in terms of the highest transmission loss measuring $10^2$.

\begin{figure}[!htb]
    \centering
    \includegraphics [scale=0.3]{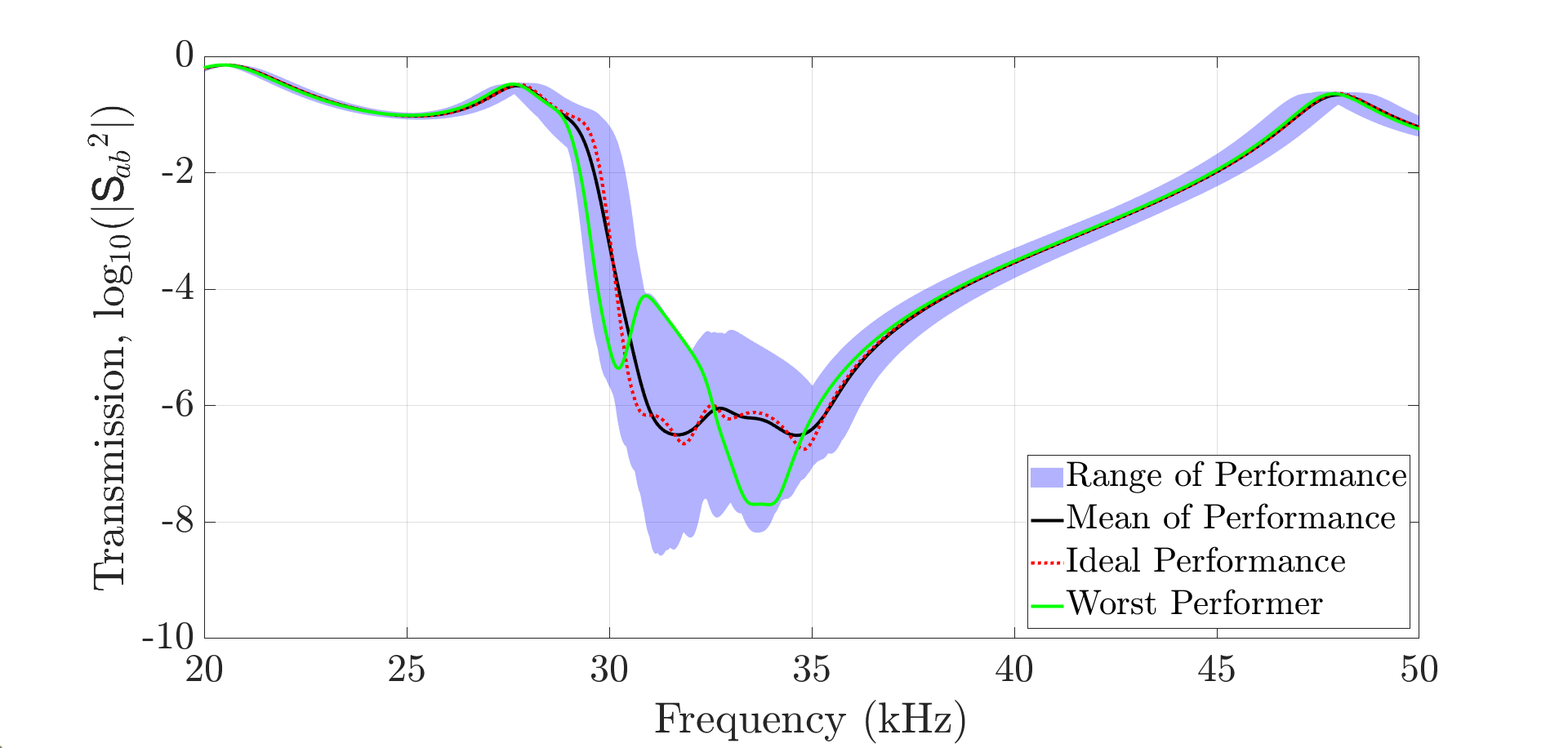}
    \caption{The graded array from Figure~\ref{square_cells} is greatly affected by $\pm 20~\mu$m randomized error, requiring improved print tolerances to avoid worst performers, or consideration of the effect of uncertainty in performance metric to be included in fitness of a specimen. $N=10000.$}
    \label{gradient_sens} 
\end{figure}

When reconstructing arrays based on the genetic algorithm solution, the resolution of the printer and its effect on the performance needs to be considered. Functionally graded structures with small steps in feature size (on the order of 20 to 50 $\mu$m) can be considered practically the same. An example of this effect is shown in Figure~\ref{gradient_sens} for the graded array identified by GA in Figure~\ref{square_cells}. Here the sample-to-sample effect is more pronounced for an applied $\pm 20~\mu$m randomized error. Each cell should be considered to have a range of performance rather than a discrete solution. 

To counter this variation, an additional objective can be added to the GA sequence which seeks to minimize an array's sensitivity to geometric error. If printer tolerance and preferred feature sizing is known, each GA function call can be expanded with a Monte Carlo sub-step that monitors the worst performers, rather than only focusing on only the strongest performers as GA typically does. This comes at non-negligible computational expense and requires powerful tools such as ROM to be feasible. The potential benefits would yield an array that has both strong attenuation performance but remains relatively insensitive to manufacturing error, greatly advancing this technology towards commercial production. This approach will be the subject of a future study.

\subsection{Improving Performance and Continuity of the Functionally Graded System}

Two major limitations on the geometric design of functionally graded arrays are apparent after optimization with ROM and genetic algorithm. The solution in Figure~\ref{wider2050_cells} contains cells for which every geometric feature is at its upper bound. This suggests that extending the upper bound would continue to produce better performing structures. The second limitation was observed in Figure~\ref{measured_sensitivity} where poor geometric accuracy, especially at sub-pixel resolution, will lead to discontinuous gradients and blurring of the transmission curve rather than discrete cell performance.

The small cell size is important to maintain as a particular strength of such low-frequency locally-resonant materials in their high figure of merit (the ratio of the wavelength in homogeneous resin to cell size). Therefore, developing larger cells is not a sustainable option and drives metamaterial design against goals identified in this field \cite{chen_review_2018,cummer_controlling_2016,liao_acoustic_2021}. An encouraging approach is to also modify the constituent materials. 3D printable resins that contain fillers possess tunable properties related to their volume fraction \cite{shah_highly_2020,sun_influences_2002,badev_photopolymerization_2011}. The bounds for ROM mass and stiffness parameters can be extended to accommodate cells with both geometric and material variation. Moreover, with a sophisticated multi-material 3D printer the materials of the web, head, and walls could be tuned individually. Increasing the contrast in mass between frame and resonator has demonstrated a significant increase in the stop band width as well as extending the applicable frequency ranges \cite{morris_expanding_2022}. 

\begin{figure*}[!htb]
  \centering
  \begin{subfigure}[htb]{0.35\textwidth}
    \centering
    \includegraphics[scale=0.2]{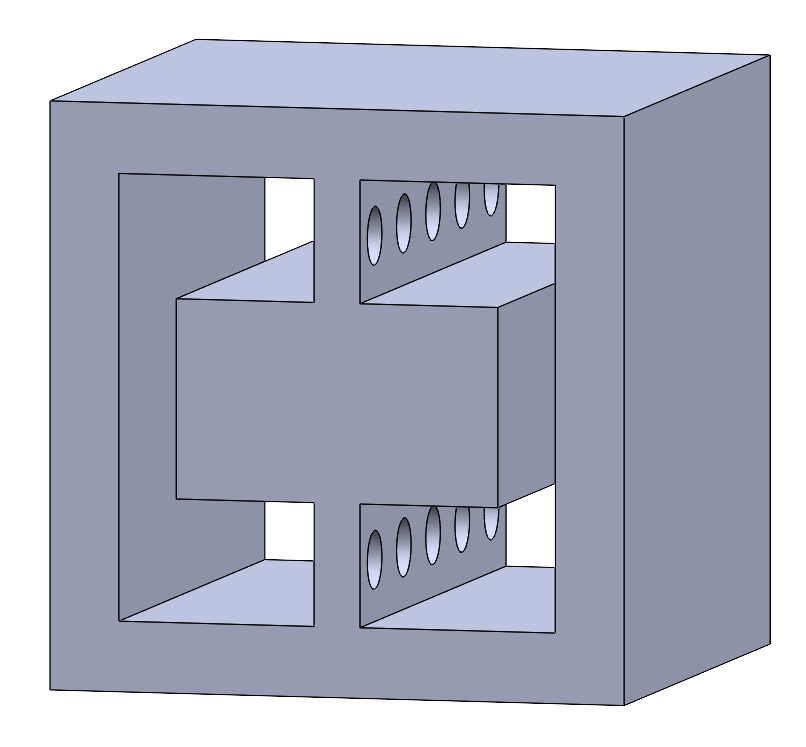}
    \caption{}
    \label{perforated_cell}
  \end{subfigure}%
  ~ 
  \begin{subfigure}[htb]{0.5\textwidth}
    \centering
    \includegraphics[scale=0.3]{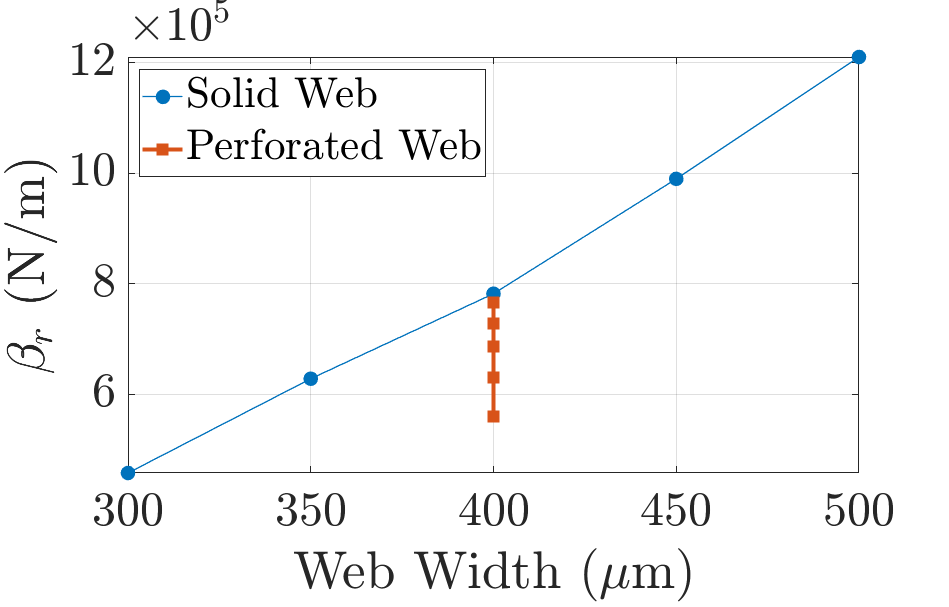}
    \caption{}
    \label{perforated}
  \end{subfigure}
  \caption{ROM parameters that require sub-pixel printing accuracy can be achieved by out-of-plane perforations, such as \subref{perforated_cell}) a series of holes in the web which \subref{perforated}) provide finer adjustment of $\beta_r$ for a fixed web width. The perforated web examples represent holes of different diameters.}
  \label{fig_perforated}
\end{figure*}

Fine tuning of ROM parameters that require sub-pixel printer accuracy could be addressed with creative design changes, such as out-of-plane perforation of sensitive features. An example is demonstrated in Figure~\ref{perforated}, where each web includes a series of 5 evenly spaced holes. In Figure~\ref{identical_performance}, it was shown that the performance metric is particularly sensitive to the web thickness, $t_w$ and its print accuracy. Variation on the scale of the pixel size ($50~\mu$m) would not enable a smoothly graded array. In ROM representation, the web width is closely associated with the resonator stiffness parameter $\beta_r$. The large shifts to $\beta_r$ as a result of $50~\mu$m steps in web width are shown in Figure~\ref{perforated}, along with values for the perforated example. The diameters of the holes in each web are stepped from 100 to 300 $\mu$m. Adjusting the perforations permits finer adjustment of $\beta_r$ while maintaining a fixed, predictable web width. The shape and spacing of the perforations can be modified to meet desired $\beta_r$ values while retaining printability in a way that decreases sensitivity to web thickness error.

\section{Conclusion}

Genetic algorithm (GA) optimization of insertion loss of a finite thickness mechanical metamaterial slab revealed several trends and design schemes that would enhance the practical performance of such systems. The approach utilizes a reduced order model (ROM) for transmission calculation, which matches well with stop band bounds and reasonably predicts band depth. Objective functions are defined in terms of ROM parameters for frame mass $M_f$, frame stiffness $\beta_f$, resonator mass $M_r$, and resonator stiffness $\beta_r$ of each cell in an array. Target performance for narrow (30 to 40 kHz) and broad (30 to 50 kHz) stop bands was achieved by minimizing integration over the stop band only. Including the frequency region outside of the stop band in the objective function produced a functionally graded structure that has strong transmission outside of a deep stop band with sharp bounds with the stop band (a square shaped transmission valley). Continuing along this trend to further extend the array (to be longer than 6 cells) may permit a smoother gradient and even more desirable performance. While minor differences exist between the reduced order (2 DOF) solution and FEM, the trends were consistent and easily justified by the substantial reduction in computational cost. A correction applied to the transfer matrix of the edge cells improved the matching between FEM and ROM analysis, though its direct use in the search and optimization requires further analysis to develop condensed formulations for such cells in contrast with the interior ones. The ROM allowed GA to solve 342210 function calls in 30 minutes, enabling optimization of a complex structure in half the time it takes to run 1 FEM model (1 hour). The ROM also enabled a study of the array's sensitivity to manufacturing tolerances. Standard deviation error of $\pm20~\mu$m introduced to the web width yields larger variation in performance while error in the head and walls was insignificant. Due to the very high speed of calculation with ROM, such sensitivities can be included in optimization objective function as well. Random error through the entire array causes a decrease in depth of $10^2$ but the mean center frequency remains on target. In some cases optimized cell geometries were pushed to the upper geometric bounds, demonstrating a need for more design freedom in order to continue improving performance. Material modification (on a part or per feature scale) would expand the ROM parameter bounds and permit additional degrees of tunability. Fine tuning of the geometry through creative design tools such as out-of-plane perforations enables control equivalent to sub-pixel resolution as well as decreased sensitivity to geometric error. 

\section*{Acknowledgements}

The authors wish to thank Dr. Farhad Pourkamali Anaraki at UMass Lowell for sharing his machine learning expertise. This research was supported by DEVCOM ARL through Cooperative Agreement W911NF-17-2-0173.

\bibliographystyle{unsrt}
\bibliography{opt_refs}
	
\end{document}